\documentclass[prd,superscriptaddress,unsortedaddress,twocolumn,showpacs,preprintnumbers,amsmath,amssymb]{revtex4}
\usepackage{graphicx}
\usepackage{bm}
\usepackage[breaklinks, colorlinks=true, pdfstartview=FitV, linkcolor=red, citecolor=blue, urlcolor=blue]{hyperref}
\usepackage{epstopdf}
\usepackage{dcolumn}
\usepackage{multirow}

\def\simge{\mathrel{\rlap{\raise 0.511ex \hbox{$>$}}{\lower 0.511ex \hbox{$\sim$}}}}
\def\simle{\mathrel{\rlap{\raise 0.511ex \hbox{$<$}}{\lower 0.511ex \hbox{$\sim$}}}}

\usepackage{amsmath}	
\begin{document}

\preprint{RBRC-976}
\title{Quark back reaction to deconfinement transition via gluon propagators}

\author{Kouji Kashiwa}
\affiliation{RIKEN/BNL Research Center, Brookhaven National Laboratory, Upton, NY-11973, USA}

\author{Yu Maezawa}
\affiliation{Physics Department, Brookhaven National Laboratory, Upton, New York 11973, USA}

\begin{abstract}
Contribution of the quark back reaction to the deconfinement phase transition is studied in the thermodynamical potential of the gluonic sector which consists of the gluon and ghost propagators calculated in the lattice QCD simulations.
Starting from QCD thermodynamic potential, we define the gluonic potential in the leading-order of the 2PI formalism, which can describe the deconfinement phase transition.
Then the gluonic potential can be written by using the microscopic characters; the gluon and ghost propagators in Landau gauge fixing.
To include effects of the quark back reaction, we calculate the gluon propagators in lattice QCD simulations with two-flavored dynamical quarks.
Fitting the lattice data by the Gribov-Stingl form and investigating the phase transition of the gluonic potential, we find that enhancement of the quark back reaction reduces the critical temperature of the deconfinement phase transition.
\end{abstract}

\maketitle


\section{Introduction}

Understanding of a nature of the deconfinement phase transition in Quantum Chromodynamics (QCD) is a long standing problem in the nuclear physics \cite{Fukushima:2010bq}.
Experimental seeking of the phase transition and quark-gluon plasma state has been done by the relativistic heavy-ion collision in RHIC and LHC~\cite{Schutz:2011zz}.
To reveal nontrivial properties of the phase transition, it is important to know thermodynamics of quarks and gluons where non-perturbative aspects of QCD play important roles.

A lattice QCD simulation is a powerful tool to deal with the non-perturbative effects of QCD and obtain computational data on the basis of the first principle calculations.
On the other hand, an effective theory is a useful approach to extract physical meanings and pictures from the numerical data.
Establishment of a reliable effective approach leads us to deep understanding of the deconfinement phase transition at finite temperature.
So far, several approaches have been proposed.
The Nambu--Jona-Lasinio (NJL) model is the most effective one to describe thermodynamics of the quark sector in terms of breaking and restoration of the chiral symmetry at finite temperature~\cite{Hatsuda:1994pi}.
However the NJL model is not approachable to describe thermodynamics of the gluonic sector which is probably essential to the deconfinement phase transition.
Recently, an extended NJL model with the Polyakov loop, so called PNJL, has been proposed \cite{Fukushima:2004,Kondo:2010ts}.
Although PNJL model can describe approximate deconfinement transition, it is not desirable model.
This is set only by macroscopic properties of the pure gauge theory
(e.g. pressure in quenched lattice QCD), since all components of the gluonic potential consists of the Polyakov loop. 
Therefore, there is no definite procedure to introduce microscopic contributions of quark excitations which couple to the gluonic sector, i.e. quark back reaction.

Microscopic description of the effective approach probably inspires us with intuitive understanding of the deconfinement phase transition.
Such an effective approach should consist of the gluon and ghost propagators which can represent fundamental degrees of freedom of the gluonic sector.
The gluon and ghost propagators also play an important role in so-called Kugo-Ojima/Gribov-Zwanziger confinement scenario \cite{Zwanziger:1994,Kugo:1995km}.
So far, the gluon and ghost propagators in the Landau gauge fixing at zero and finite temperature have been studied in the lattice simulations, the Dyson-Schwinger (DS) equation approach, and the functional renormalization group (FRG) approach~\cite{Vandersickel:2012tz,Fischer:2010fx,Aouane:2012}.
Recently, FRG approach has found that the deconfinement transition can be described by the Polyakov loop via the Landau-gauged gluon and ghost propagators~\cite{Pawlowski:2004,Braun:2007bx}, and the quark back reaction contributes to critical temperature of the phase transition~\cite{Pawlowski:2010ht}.
The gluon and ghost propagators in FRG approach are well described in the pure-gauge limit.
However, contributions arising from the quark back reaction are expressed by the quark-meson model which starts from the bare mesonic potential at ultra-violet scale and represents the quantum fluctuation via the meson propagations.
Therefore it is uncertain to describe appropriate quark excitations based on the first principle of QCD.

To establish sound effective approach from the microscopic degrees of freedom, it has been proposed that the gluonic potential is formulated by starting from the pure gauge theory based on the leading contributions of the 2PI formalism \cite{Fukushima:2012}.
This approach enables us to deal with the propagators obtained in the lattice simulations which include all information of the non-perturbative aspects of QCD.
They have found that several physical quantities (e.g. the critical temperature, equation of states, etc.) can be reproduced via the approach with input of the propagators determined in pure-gauge lattice simulations.

In this paper, we develop the previous approach to joining the gluon propagators determined in the lattice QCD simulations with two-flavored dynamical (sea) quarks at finite temperature and investigate the effect of the quark back reaction to the deconfinement phase transition.
At finite temperature, the four-dimensional transverse gluon propagator can be decomposed into the tree-dimensional longitudinal and transverse components.
We investigate temperature dependence of the propagators in lattice QCD simulations.
By using these propagators with the Gribov-Stingl form, we calculate the critical temperature via the effective potential and investigate the effect of the quark back reaction to the phase transition of the gluonic sector.

This paper is organized as follows.
In the next section, we introduce the formalism of the effective approach and calculate the thermodynamical potential of the gluonic sector.
The lattice QCD formalism and simulations are shown in Sec. III and the critical temperature calculated by the gluonic potential is shown in Sec. IV.
The paper is summarized in Sec. V

\section{Gluon and ghost effective potential}

First of all, let us briefly explain formulation of the effective potential of the gluonic sector.
In general, the effective approach begins to distinguish the QCD thermodynamical potential to matter and gluonic sectors,
\begin{align}
\Omega_{\rm QCD} &\rightarrow \Omega_\mathrm{mat} + {\cal U}.
\end{align}
One of the descriptions of the matter part is, for instance, NJL model.
Mixture between $\Omega_\mathrm{mat}$ and ${\cal U}$ is induced by a classical field $\phi$ which appears in the covariant derivative of $\Omega_\mathrm{mat}$.
We can define it as the temporal components of the gluon field:
$\phi = \phi_3 \lambda_3 + \phi_8 \lambda_8$
where $\lambda_3$ and $\lambda_8$ are diagonal components of Gell-Mann matrices.
At zero chemical potential, only $\lambda_3$ path becomes relevant, and we can rewrite it only with one angular parameter $\theta$:
$\phi = 2 \pi \times \mathrm{diag} (\theta,-\theta,0)$.
Then the Polyakov loop can be written as
\begin{align}
\Phi &= \frac{1}{N_\mathrm{c}} \mathrm{tr_c} e^{i\phi} =
 \frac{1}{N_\mathrm{c}} [ 1 + 2 \cos(2\pi \theta) ].
\end{align}

In this paper, we focus only on the thermodynamical potential of the gluonic sector and construct it by the gluon and ghost propagators in the Landau gauge fixing.
The effective potential of the gluonic sector can be evaluated by the leading-order contribution of the 2PI formalism in the pure-gauge theory~\cite{Fukushima:2012},
\begin{align}
{\cal U} =& - \frac{1}{2} \mathrm{tr} \ln D_A^{-1} + \mathrm{tr} \ln D_C^{-1} ,
\label{TP-or}
\end{align}
where ``tr'' acts on all indices.
Quantum fluctuation of $\phi$ and spatial gluon fields have already been integrated out and contributes via the gluon $D_A$ and ghost $D_C$ propagators.
$D_A$ can be expressed by the four-dimensional (4d) transverse ($T_{\mu \nu}$) and longitudinal ($L_{\mu\nu}$ ) projection tensor,
\begin{align}
D_A (p^2) &= T_{\mu \nu} D_{(T)}(p^2) + L_{\mu \nu} D_{(L)} (p^2)
,
\end{align}
with 4d transverse ($D_{(T)}$) and longitudinal ($D_{(L)}$) components.
Since the Lorentz symmetry is explicitly broken at finite temperature, the 4d transverse gluon propagator is decomposed into the three-dimensional (3d) transverse and longitudinal components:
\begin{align}
T_{\mu \nu} D_{(T)} (p^2) &= P^T_{\mu \nu} D_T + P^L_{\mu \nu} D_L.
\end{align}
where we impose the Landau-gauge fixing.
Then the gluon propagators can be defined by the gluon fields as
\begin{align}
D_T (p) 
& = \frac{1}{2N_g}
    \left\langle A_i^a A_i^a - \frac{p_4^2}{{\vec p}^2} A_4^a A_4^a \right\rangle,
\label{DT}\\
D_L (p) 
& = \frac{1}{N_g} 
    \left( 1 + \frac{p_4^2}{{\vec p}^2} \right) 
    \left\langle A_4^a A_4^a \right\rangle,
\label{DL}
\end{align}
where $N_g = N_c^2 -1$.
In the following, we impose the zero Matsubara-frequency on the propagators, i.e. $p_4 = 0$.
The effective potential (\ref{TP-or}) becomes
\begin{align}
{\cal U} =& - \frac{1}{2} \Bigl[ 
              \mathrm{tr} \ln (D_{(L)}^{-1}) + \mathrm{tr} \ln (D_C^{-1})
\nonumber\\
          & +\mathrm{tr} \ln (D_L^{-1}) + 2 \mathrm{tr} \ln (D_T^{-1}) \Bigr],
\label{TP-or2}
\end{align}
where traces of the Lorentz indices are already done.

Since analytic forms of the propagators help us to evaluate the effective potential in practice, we assume the ghost propagator as
\begin{align}
D_C(p^2) &= \frac{p^2 + d_C^{-2} }{(p^2)^2},
\label{GhF}
\end{align}
and the gluon propagator as the ``Gribov-Stingl'' form:
\begin{align}
D (p^2) &= \frac{c^2 d^2 (p^2 + d^{-2})}{(p^2+r^2)^2},
\label{GLF}
\end{align}
where subscripts $L$ and $T$ for the gluon propagator and parameters are omitted.
The latter form can be derived in a low-energy effective reduction of the
Yang-Mills theory~\cite{Kondo:2011ab}.  
For the ghost propagator, we utilize results of $D_C$ obtained in Ref.~\cite{Fukushima:2012} by fitting the data in pure-gauge lattice simulations~\cite{Aouane:2012}, since the ghost propagator is not sensitive to the dynamical quarks due to lack of the quark-ghost vertex \cite{Ilgenfritz:2007iq}.

By substituting Eqs.~(\ref{GhF}) and (\ref{GLF}), the effective potential (\ref{TP-or2}) becomes
\begin{align}
{\cal U}
&= \Bigl( \frac{1}{2} - 2 \Bigr) {\cal V}_\mathrm{pert}
- \frac{1}{2} \Bigl( {\cal V}_A^{(L)}(d_L^{-2}) - 2 {\cal V}_A^{(L)}(r_L^2) \Bigr) 
\nonumber\\
&- \frac{2}{2} \Bigl( {\cal V}_A^{(T)}(d_T^{-2}) - 2 {\cal V}_A^{(T)}(r_T^2) \Bigr)  + {\cal V}_C (d_C^{-2}),
\label{GGP}
\end{align}
where ${\cal V}_\mathrm{pert}$,${\cal V}_A^{(L)}$, ${\cal V}_A^{(T)}$ and ${\cal V}_C$ are the perturbative one-loop~\cite{Gross:1981,Weiss:1981}, 3d longitudinal gluon, 3d transverse gluon and ghost effective potentials, respectively. 
The gluon and ghost effective potentials can be expressed as
\begin{align}
{\cal V} (m^2;\theta) 
&= - T \mathrm{tr}_c \int \frac{dp}{2 \pi^2} 
     \ln \Bigl( 1 - L_8 e^{ -\beta \sqrt{p^2+m^2} } \Bigr),
\label{gs}
\end{align}
where $\mathrm{tr}_c$ is the adjoint color trace and $L_8$ is the Polyakov-loop operator in the adjoint representation $(L_8)_{ab} = 2 \mathrm{tr}_c (t_a e^{i\phi} t_b e^{i\phi})$. 
Equation (\ref{GGP}) means that, if there are the parameters $(d_L, r_L, d_T, ...)$ determined in the lattice QCD simulations, one can calculate the gluonic potential from microscopic points of view:
The gluon and ghost propagators which basically manifest all non-perturbative aspects of QCD including the quark back reaction.

It is worth to mention that in the derivation of PNJL model, the spatial gluons are expressed by the Polyakov loop via the strong coupling expansion and the quantum fluctuation of the temporal gluons is suppressed due to an application of the classical Haar measure \cite{Fukushima:2004}.
This means that there is no microscopic component (like a propagator) in the gluonic sector of PNJL model, and it is impossible to deal with the quark back reaction based on a definite manner.

\section{Gluon propagators on lattice}

In order to investigate effects of the quark back reaction to the phase transition of the gluonic sector, it is necessary to calculate the gluon propagators in the lattice QCD simulations with dynamical quarks.
We utilize gauge configurations generated by WHOT-QCD Collaboration on $N_s^3 \times N_t = 16^3 \times 4$ lattice with a renormalization-group improved gauge action $S_g$ and a clover improved Wilson quark action with two flavors $S_q$:
\begin{align}
S_g &= -\beta \sum_x \Bigl( c_0 \sum_{\mu < \nu} W_{\mu\nu}^{1\times1} (x) + c_1 \sum_{\mu \ne \nu} W_{\mu\nu}^{1\times2} (x) \Bigl),\\
S_q &= \sum_{f=1,2}\sum_{x,y} \bar{q}_x^f D_{xy} q_y^f ,
\end{align}
where $c_1=-0.331$, $c_0 = 1-8c_1$ and $W_{\mu\nu}^{n \times m}$ is a $n \times m$ shaped Wilson loop.
The quark matrix is
\begin{align}
&D_{xy} = \delta_{xy} - K \sum_\mu [ (1-\gamma_\mu) U_{x,\mu} \delta_{x+\hat{\mu},y} \nonumber \\
&                                     +(1+\gamma_\mu) U_{x,\mu}^\dagger \delta_{x,y+\hat{\mu}} ] 
                      - \delta_{xy} c_{SW} K \sum_{\mu<\nu} \sigma_{\mu\nu} F_{\mu\nu} ,
\end{align}
where $K$ is the hopping parameter, $F_{\mu\nu}$ is the lattice field strength and
 $c_{SW}$ is the clover coefficient determined in the one-loop perturbation theory:
$c_{SW} = (1 - 0.8412 \beta^{-1})^{-3/4}$.
The simulations have been performed along the line of constant physics at $m_{\rm PS}/m_{\rm V} = 0.65$ (ratio of the pseudo-scalar and the vector meson masses at zero temperature).
Details of the lines of constant physics and the phase diagram with the same lattice actions are given in Refs. \cite{Maezawa:2007fc,Ejiri:2009hq}.
For the investigation of the phase transition, we calculate the propagators in the temperature range of $0.8 \simle T/T_{\rm pc} \simle 1.4$, where $T_{\rm pc}$ is the pseudo-critical temperature for the transition from the hadronic phase to the quark-gluon plasma phase in $N_f=2$ QCD.

The purpose of the lattice simulations is to evaluate the gluon propagators (\ref{DT}) and (\ref{DL}).
The gauge field on the lattice can be defined by the link valuable as 
\begin{align}
A_{\mu}(x) &\equiv \frac{1}{ 2 i a} \left[ U_{x,\mu} - U^\dag_{x,\mu} \right] ,
\end{align}
where $a$ is a lattice spacing and we assume the isotropic lattice.
Imposing the Landau gauge fixing on the lattice configurations, we define the zero-frequency ($p_4 = 0$) gluon propagators in the momentum space,
\begin{align}
D_T (p) &= \frac{a^4}{2N_g} \sum_{i=1}^3 \sum_{x}  {\rm tr} \left[ A_i (x) A_i (0) \right] e^{-2 \pi i k_j x_j / N_s } ,\\
D_L (p) &= \frac{a^4}{N_g}               \sum_{x}  {\rm tr} \left[ A_4 (x) A_4 (0) \right] e^{-2 \pi i k_j x_j / N_s } ,
\end{align}
where $k_j = 0, 1,\cdots ,N_s-1$ and the physical momentum is 
\begin{align}
p = \frac{2}{a} \sqrt{ \sum_i \sin^2 \left( \frac{2 \pi k_i}{N_s} \right) }.
\end{align}
Notice that the transeverse condition $p_i A_i =0$ is imposed on $D_T(p)$.
We measure the propagators with 5000 trajectories and estimate statistical errors by the jackknife method with bin size of 100 trajectories.

\begin{figure}[tb]
\begin{center}
 \includegraphics[width=0.5\textwidth]{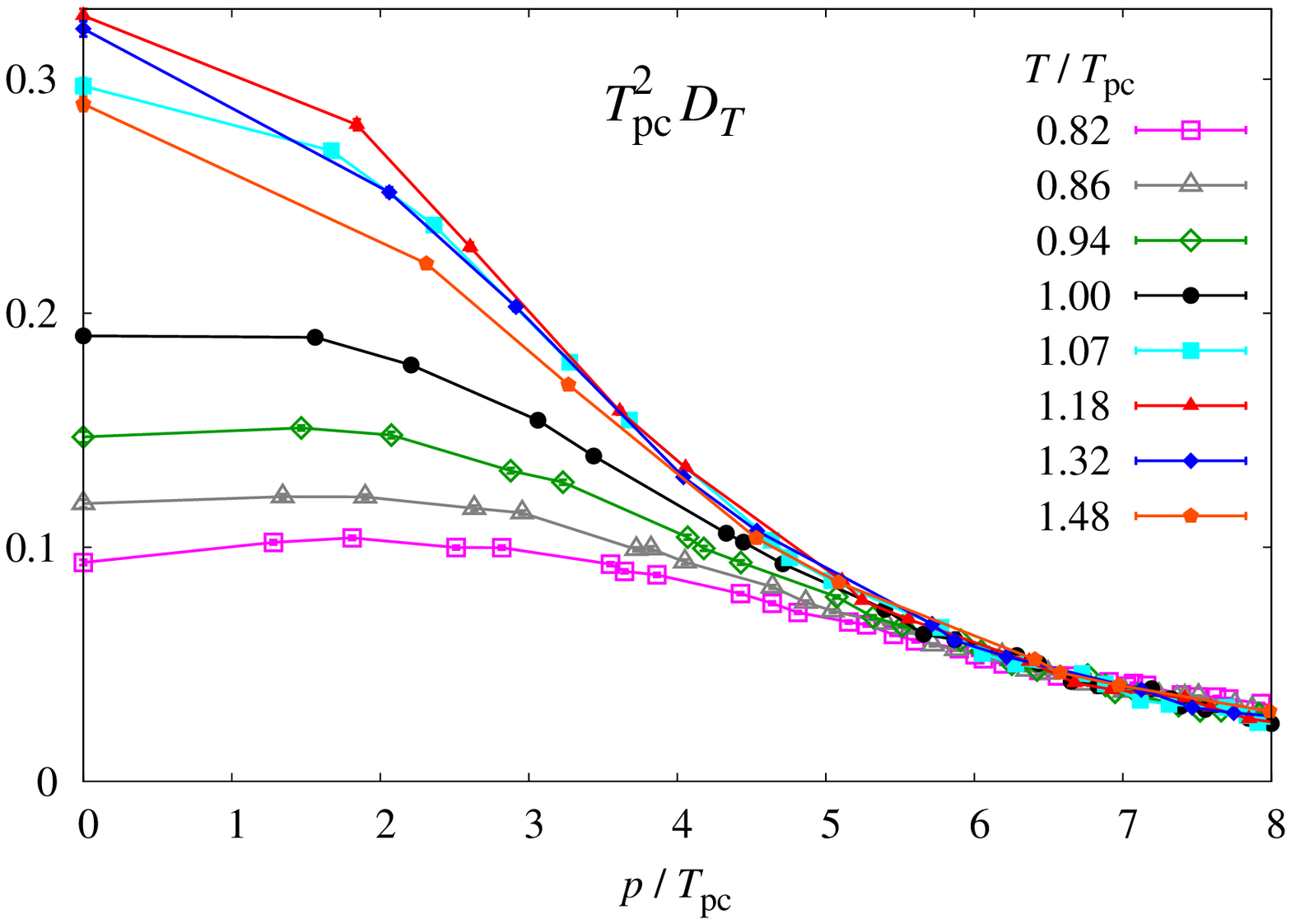}
 \includegraphics[width=0.5\textwidth]{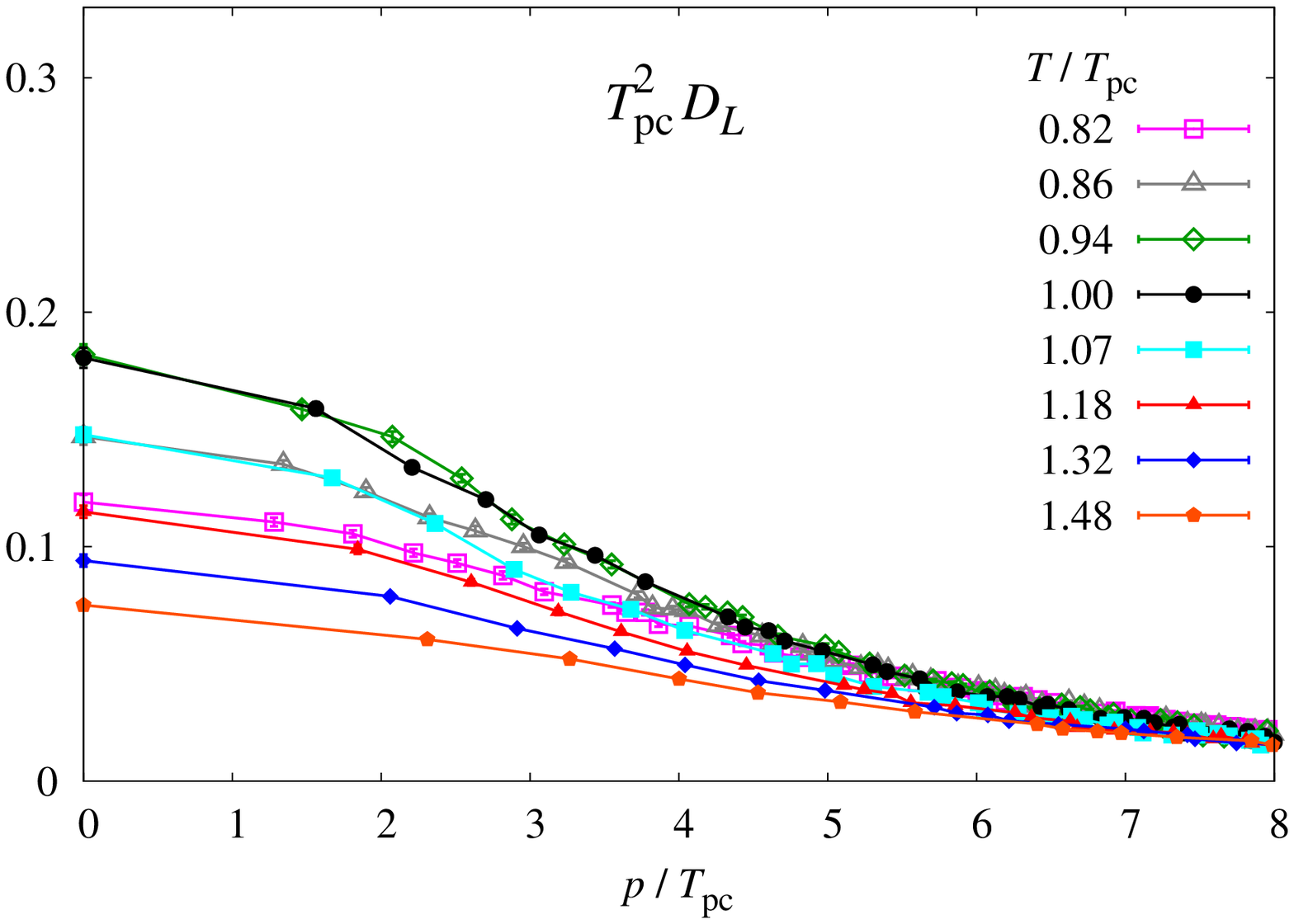}
\end{center}
\caption{Results of the transverse (top) and longitudinal (bottom) gluon propagators in momentum space for several temperature.
 A physical scale is normalized by $T_{\rm pc}$.}
\label{Fig:T-GP-L-H}
\end{figure}

Figure \ref{Fig:T-GP-L-H} shows temperature dependence of $D_T$ (top) and $D_L$ (bottom) as a function of momentum, where 
 a physical scale is normalized by $T_{\rm pc} \sim 180$ MeV.
We find that there is significant temperature dependence at infra-red region (i.e. small $p$).
On the other hand, the propagators converge at ultra-violet region where the physics is insensitive to temperature and probably dominated by perturbative descriptions. 
When temperature increases, magnitude of both propagators increases at $T<T_{\rm pc}$ and shows enhancement at $T \sim T_{\rm pc}$ for $D_L$ and $T \sim 1.2T_{\rm pc}$ for $D_T$ and decreases subsequently.
In the case of pure gauge theory, the gluon propagators are almost insensitive to temperature below $T_{\rm c}$, whereas these monotonically decrease above $T_{\rm c}$ \cite{Aouane:2012}.
Therefore the strong sensitivity to temperature below $T_{\rm pc}$ in our case implies that there are a large number of configurations which can couple to the gluon fields even at $T<T_{\rm pc}$ where the gluon confinement is slightly broken by the quark back reaction.
Since the number of such configurations would be proportional to magnitude of the gluon propagators, the quark back reaction enhances with temperature increases at $T \simle T_{\rm pc}$.

Furthermore, the enhancement of the propagators in momentum space means strong correlations between gluons in real space.
In general, the longitudinal propagator is dominated by the electric gluon sector which is sensitive to the deconfinement phase transition, e.g. Polyakov loop which consists of a sum of the longitudinal gluons.
It is therefore natural that the correlation of longitudinal gluons becomes strong at $T \sim T_{\rm pc}$.
On the other hand, the transverse propagator is dominated by the magnetic gluon sector which is probably insensitive to the phase transition.
For instance, the confinement of spatial Wilson loop, characterized by the spatial string tension, does not disappear even above $T_{\rm pc}$ \cite{Maezawa:2007fc}.
However the enhancement of $D_T$ at $T \sim 1.2 T_{\rm pc}$ implies that the magnetic sector may also play an important role in the deconfinement phase transition.

\section{Quark back reaction to effective potential}

\begin{table*}[t]
  \begin{minipage}{.65\textwidth}
     \caption{Summary of fits with $\chi^2/N_{\rm DF}$ for several temperatures.
 We take two types of fit ranges for each temperature.}
 \vspace{-5mm}
   \begin{center}
 {\renewcommand{\arraystretch}{1.0} \tabcolsep = 3mm
 \newcolumntype{.}{D{.}{.}{2}}
 \begin{tabular}{cc......}
 \hline\hline
 \multicolumn{1}{c}{$T/T_{\rm pc}$} & 
 \multicolumn{1}{c}{$p/T_{\rm pc}|_{\rm min}$} & 
 \multicolumn{1}{c}{$d_L^{-2}$}&
 \multicolumn{1}{c}{$r_L^2$}&
 \multicolumn{1}{c}{$\chi^2/N_{\rm DF}$}&
 \multicolumn{1}{c}{$d_T^{-2}$}&
 \multicolumn{1}{c}{$r_T^2$}&
 \multicolumn{1}{c}{$\chi^2/N_{\rm DF}$}
 \\
 \hline
\multirow{2}{*}{0.82} & 3.64 & 1.24(7) & 0.52(1) & 0.86 & 2.3(1) & 0.827(9) & 3.47 \\
                      & 4.64 & 1.60(8) & 0.534(8)& 1.13 & 2.1(1) & 0.696(7) & 1.41 \\
\hline
\multirow{2}{*}{0.86} & 3.82 & 2.1(2)  & 0.49(1) & 1.73 & 1.82(5)& 0.592(8) & 3.51 \\
                      & 4.86 & 2.8(2)  & 0.46(1) & 0.14 & 1.49(8)& 0.460(8) & 1.03 \\
\hline
\multirow{2}{*}{0.94} & 4.18 & 3.5(3)  & 0.47(1) & 0.87 & 0.84(2)& 0.287(8) & 12.09 \\
                      & 5.32 & 4.3(3)  & 0.442(1)& 0.13 & 0.75(1)& 0.145(7) & 2.90 \\
\hline
\multirow{2}{*}{1.00} & 4.44 & 4.8(4)  & 0.50(2) & 1.70 & 0.61(1)& 0.114(8) & 11.28 \\
                      & 5.66 & 3.6(2)  & 0.33(1) & 0.76 & 15.0(1.4)& 0.59(1)  & 0.43 \\
\hline\hline
  \end{tabular}}    
  \end{center}
    \label{tab:para}
  \end{minipage}
\begin{minipage}{.1\textwidth}
\end{minipage}
 \begin{minipage}{.3\textwidth}
    \caption{Results of $T_c'$ evaluated from thermodynamic potential of the gluonic sector at several $T/T_{\rm pc}$.
  The first (second) parenthesis indicates statistical (systematic) errors.}
    \begin{center}
 {\renewcommand{\arraystretch}{1.2} \tabcolsep = 3mm
 \begin{tabular}{cc}
\hline\hline
 $T/T_{\rm pc}$ & $T_c'$ [GeV]\\
 \hline
0.82 & 0.332(20)(10) \\
0.86 & 0.291(26)(11) \\
0.94 & 0.233(19)(24) \\
\hline\hline
  \end{tabular}}
    \end{center}
    \label{Table:fit_each}
  \end{minipage}
\end{table*}

In this section, we calculate the critical temperature from the thermodynamic potential of the gluonic sector (\ref{GGP}) by using the gluon and ghost propagators on the lattice, and investigate effects of the quark back reaction to the phase transition.
To input the lattice results to the thermodynamic potential, we fit the gluon propagators on the lattice shown in Fig.~\ref{Fig:T-GP-L-H} by the ``Gribov-Stingl'' form (\ref{GLF}).
Since, in general, a finite size effect becomes serious at small $p$ \cite{Aouane:2012}, we cut the fit range at some momentum, i.e. taking the fit range at $p \ge (p/T_{\rm pc})_{\rm min}$.
To consider the fit range dependence, we take two values of $(p/T_{\rm pc})_{\rm min}$.
We also use only the data at $p=(p,0,0)$ to reduce artifacts of rotational symmetry breaking on the lattice.
The fit is performed by minimizing $\chi^2$ and results are summarized in Tab.~\ref{tab:para}.

Results show that at $T/T_{\rm pc}=0.83$ and $0.86$ the fit provides good $\chi^2$s, whereas it does not work well when temperature increases.
The fit range dependence are less than 20 \% at low temperature, whereas it becomes huge at $T/T_{\rm pc} = 1.00$.
Therefore, we restrict applicable temperature of the ``Gribov-Stingl'' form and our effective approach within $T/T_\mathrm{pc} \le 0.94$.

Then the critical temperature of the effective potential $T'_c$ can be obtained by the following steps:
\\
1. Determine the parameters $d_{L,T}$ and $r_{L,T}$ at each \\
~~~~$T/T_\mathrm{pc}$.\\
2. Minimize the effective potential (\ref{GGP}) with respect to \\
~~~~the classical field $\phi$.\\
3. Determine $T_c'$ at the point that the Polyakov-loop \\
~~~~becomes non-zero value.\\
Notice that $T_c'$ does not correspond to $T_\mathrm{pc}$ on the lattice because it is obtained by the {\it purely} gluonic sector $(\ref{GGP})$.
Qualitative shape of the effective potential (\ref{GGP}) is the same with that of the previous study using the gluon propagators in pure gauge theory \cite{Fukushima:2012}.
Therefore it exhibits the first order phase transition.
Contribution of the quark back reaction appears as a shift of the characteristic scale $T_c'$ in the effective approach.

Table~\ref{Table:fit_each} summarizes results of $T_c'$ in GeV unit for several $T/T_{\rm pc}$.
We estimate $T_c'$ as an average of two fit ranges shown in 
 Tab.~\ref{tab:para} and systematic errors due to the fit range dependence as difference from the averaged value, which indicate the second parenthesis in the Table.
We can see that $T_c'$ is sensitive to $T/T_{\rm pc}$ and decreases with temperature increasing at $T<T_{\rm pc}$.
Since, as mentioned above, the number of lattice configurations which couple to the gluon fields increases with temperature increases at below $T_{\rm pc}$, the decrease of $T_c'$ is caused by the enhancement of the quark back reaction via the gluon propagators.

\section{Summary}

We have investigated effects of the quark back reaction to the gluonic thermodynamical potential which consists of the gluon and ghost propagators.
We have calculated the gluon propagators of the 3-dimensional longitudinal and traverse sectors in lattice QCD simulations with two-flavor dynamical quarks at finite temperature.
The simulations have been performed on $16^3 \times 4$ lattice at the temperature range of $0.8 \simle T/T_{\rm pc} \simle 1.4$, where $T_{\rm pc}$ is the pseudo-critical temperature on the lattice.
 
Results of the lattice simulations have shown that magnitude of the gluon propagators increases at $T<T_{\rm pc}$.
This can be regarded as contributions of the quark back reaction since the gluon propagators are almost insensitive below the critical temperature in the case of the pure gauge theory \cite{Aouane:2012}.
This implies that there are many configurations which can couple to the gluon fields even at $T<T_{\rm pc}$ since the gluon confinement is slightly broken by the quark back reaction.
The longitudinal (transverse) propagator has shown enhancement at $T \sim T_{\rm pc}$ $(1.2 T_{\rm pc})$ and then both decrease when temperature increases.
This means that not only the longitudinal propagator but also the transverse one should be of importance to describe the deconfinement phase transition since the enhancement of the propagators in momentum space indicates strong correlations in real space which may relate to divergence of the correlation length in the QCD phase transition. 

In order to estimate contributions of the quark back reaction to the deconfinement phase transition, we fit the propagators by ``Gribov-Stingl'' form and input these to gluonic potential obtained by the leading-order of the 2PI formalism in the pure-gauge theory \cite{Fukushima:2012}.
The critical temperature estimated in the gluonic potential decreases with temperature of the gluon propagators increasing at $T \simle T_{\rm pc}$ 
This means that the enhancement of the quark back reaction reduces the critical temperature of the gluonic sector.

In this paper, we discuss qualitative contributions of the quark back reaction to the gluonic potential.
It has however been found that the strong sensitivity to temperature even below $T_{\rm pc}$ appears in two-flavor QCD, in contrast to the pure gauge theory.
This implies that it is indispensable to introduce a temperature-dependent Gribov-Stingl form for a precise calculation of the critical temperature in the gluonic sector $T_c'$.
In general, the temperature dependent parameters in the Gribov-Stingl form make the analytic calculation of the gluon potential (\ref{gs}) difficult.
Then one needs to numerically calculate summations of the Matsubara frequency and subtract divergence of the propagators by using the same operators at zero temperature on the lattice, which require huge numerical cost.
Furthermore, to obtain abundant and detailed information of momentum dependence on the lattice, simulations on large lattice are important.
In particular, it is known that the finite size effect of the gluon propagators becomes serious at infrared regions.
Although we omit the data at small momentum, to draw a definite conclusion about the quark back reaction we need careful investigation with larger lattice size.
We leave these to our future work.
Establishment of reliable effective potential of the gluonic sector combined with the matter potential $\Omega_{\rm mat}$ will lead us to the understanding of the phase structure of QCD at finite temperature and density.

{\it Acknowledgements:}
We thank WHOT-QCD Collaboration for providing us with the gauge configurations.
The authors thank Kenji Fukushima for useful comments.
K.K. is supported by RIKEN Special Postdoctoral Researchers Program.
The numerical simulations were performed in the RIKEN Integrated Cluster of Clusters (RICC).

\bibliography{paper}

\end{document}